\pdfoutput=1
\documentclass[aps,prx,twocolumn,longbibliography,superscriptaddress]{revtex4-2}
\usepackage{amsmath}

\usepackage{bm}
\usepackage{bbold}
\usepackage{amssymb}
\usepackage{graphicx}
\usepackage[ hidelinks = true, colorlinks = true, citecolor = blue, urlcolor = blue, linkcolor = blue]{hyperref}
\usepackage{verbatim}
\usepackage{color}
\usepackage[capitalise]{cleveref}
\usepackage[normalem]{ulem}
\usepackage{dsfont}

\allowdisplaybreaks[3]

\usepackage{dsfont}

\usepackage{accents}
\newlength{\dhatheight}

\begin{document}

\title{Thermoelectric cooling of a finite reservoir coupled to a quantum dot}

\author{Stephanie Matern}
\affiliation{Pitaevskii BEC Center, CNR-INO and Dipartimento di Fisica, Universit\`a di Trento, 38123 Trento, Italy}
\affiliation{NanoLund and Solid State Physics,  Lund University,  Box 118,  22100 Lund, Sweden}
\author{Saulo V. Moreira }
\affiliation{School of Physics, Trinity College Dublin, Dublin 2, D02 K8N4, Ireland}
\affiliation{NanoLund and Mathematical Physics,  Lund University,  Box 118,  22100 Lund, Sweden}
\author{Peter Samuelsson}
\affiliation{NanoLund and Mathematical Physics,  Lund University,  Box 118,  22100 Lund, Sweden}
\author{Martin Leijnse}
\affiliation{NanoLund and Solid State Physics,  Lund University,  Box 118,  22100 Lund, Sweden}

\date{\today}

\begin{abstract}

 We investigate non-equilibrium transport of charge and heat through an interacting quantum dot coupled to a finite electron reservoir. Both the quantum dot and the finite reservoir are coupled to conventional electric contacts, i.e., infinite electron reservoirs, between which a bias voltage can be applied. We develop a phenomenological description of the system, combining a rate equation for transport through the quantum dot with standard expressions for bulk transport between the finite and infinite reservoirs. The finite reservoir is assumed to be in a quasi-equilibrium state with time-dependent chemical potential and temperature which we solve for self-consistently. We show that the finite reservoir can have a large impact on the stationary state transport properties, including a shift and broadening of the Coulomb diamond edges. We also demonstrate that there is a region around the conductance lines where a heat current flows out of the finite reservoir. Our results reveal the dependence of the temperature that can be reached by this thermoelectric cooling on the system parameters, in particular the coupling between the  finite and infinite reservoirs and additional heat currents induced by electron-phonon couplings, and can thus serve as a guide to experiments on quantum dot-enabled thermoelectric cooling of finite electron reservoirs. Finally, we study the full dynamics of the system, with a particular focus on the timescales involved in the thermoelectric cooling.
\end{abstract}

\maketitle

\section{Introduction}

Experimental advances in the fabrication of nanoscale electronic devices have made it possible to employ measurements of non-equilibrium transport to deduce the properties of a (quantum) system. Such devices include a wide range of systems, e.g.,  molecular junctions, carbon nanotubes, graphene structures and quantum dots (QDs)\cite{Gehring2019,Laird2015,Bischoff2015,Wiel2002,Hanson2007}.
They are a fundamental building block for applications such as transistors, (thermal) diodes, nanoscale heat engines, and sensors \cite{DiezPerez2009,Perrin2015,Benenti2017,Jaliel2019,Malik2022,Tesser2022,Blasi2023,Josefsson2018}.   
From a theoretical point of view, the description of open quantum systems out of equilibrium remains a challenging task~\cite{Landi2022}. In non-equilibrium transport scenarios, it is common to assume that the system is in contact with infinite reservoirs at different temperatures and chemical potentials. Although this is a suitable description for a number
of experimental situations, accounting for the finite aspect of reservoirs is needed in many nanoscale, non-equilibrium scenarios \cite{Reimann2008}.
Various experiments on, e.g., superconducting~\cite{Houck2012, Schmidt2013} or ultracold atom platforms~\cite{ Brantut2015} demonstrate intricate control over the non-equilibrium properties of the system, including access to the state of the reservoir~\cite{Krinner2017,Lebrat2018}.

Therefore, finite reservoirs have 
recently attracted increased attention in theoretical investigations of non-equilibrium phenomena.
Efforts to understand how finite reservoirs affect transport properties have involved the development of more appropriate transport models~\cite{Zwolak2020,Trushechkin2022}, encompassing relaxation processes in QD systems~\cite{Schaller2014} and ultracold atoms~\cite{Gallego-Marcos2014}, many-particle transport~\cite{Amato2020} and the full system dynamics~\cite{Ajisaka2013,Riera2021}.
Further theoretical works in the literature address temperature fluctuations in metallic islands~\cite{Berg2015} and their thermodynamic implications~\cite{Moreira2023}. 
Considering a finite reservoir as a heat source for an engine, the optimization of such a heat engine's efficiency has been investigated~\cite{Wang2016, Pozas-Kerstjens2018, Strasberg2021, Yuan2022}.

Particle transport also induces the exchange of energy and heat between different parts of a system. It is natural to consider thermal properties and thermoelectric behavior of finite reservoirs, which are the subject of ongoing interest~\cite{Brander2015}. 
Generally, thermoelectric effects describe a system's or material's ability to convert a temperature difference to an electric voltage (or vice-versa). 
A high thermoelectric efficiency results from strong energy asymmetries in electron transport properties~\cite{Mahan1996,Humphrey2005,Benenti2017}, which result from sharp resonances in meso- and nanoscale systems ~\cite{Giazotto2006,Dubi2011,Bevilacqua2022}.
They readily occur in QD devices, promoting QDs to an ideal platform for experimental studies of thermoelectric behavior in quantum devices, see e.g.~\cite{Staring1993,Scheibner2007,Svensson2012,Svilans2016,Josefsson2018,Danial2022}.
Proposals to boost the thermoelectric response of quantum systems further include utilizing non-linear \cite{Whitney2013,Markos2021} and coherence effects \cite{Karlstrom2011,Trocha2012},  or topological states \cite{Hajiloo2020}.

An immediate use of thermoelectric properties is Peltier cooling, where a bias voltage can be used to force heat to flow from a cold to a hot reservoir.  Such electronic cooling has  been realized experimentally using QD systems~\cite{Prance2009},  molecular junctions \cite{Cui2018},   normal metal-insulator-superconductor junctions \cite{Muhonen2009,Nguyen2013}
and has been investigated in cold atomic gases~\cite{Brantut2013, Grenier2016}.
In this context, given that any physical device or system being refrigerated is necessarily of a finite size, it is essential to understand transport and thermoelectric behaviors in non-equilibrium settings with finite reservoirs.

In this paper we set out to investigate transport and thermoelectric properties of a system consisting of  a finite electron reservoir in contact with a QD. 
We assume a quasi-equilibrium state of the finite reservoir~\cite{Schaller2014,Amato2020}, which can be described by a local equilibrium distribution.  To allow for stationary-state non-equilibrium transport, the system is additionally coupled to two infinite electron reservoirs, see \cref{fig:setup}. When there is an applied voltage and/or a temperature difference between the infinite reservoirs, we self-consistently solve for the chemical potential and temperature of the finite reservoir together with the exchanged heat and charge currents, which also means that these quantities become time-dependent.
We investigate the transport properties in the stationary state, as well as the full dynamics.
We find that the finite reservoir modifies the stationary state properties considerably compared to treating all reservoirs as infinite. 
Additionally, we discover a parameter regime where heat is carried out of the finite reservoir, resulting in a stationary-state temperature of the finite reservoir \emph{lower} than the temperature of the infinite reservoirs.

The paper is organized as follows. In \cref{sec:model}, the system is introduced, and its non-equilibrium dynamics is modeled using a combination of a rate equation for the QD and standard expressions for bulk systems  for transport  between the finite and infinite reservoirs. In \cref{sec:transport}, we solve for the non-equilibrium stationary state and analyse the impact of the finite reservoir on the transport through the QD. The thermoelectric refrigeration of the finite reservoir is discussed in \cref{sec:cooling}. Finally, we investigate the system's full dynamics in \cref{sec:transient}.

\section{Model} \label{sec:model}

\begin{figure}[t]
    \centering
    \includegraphics[width = \linewidth]{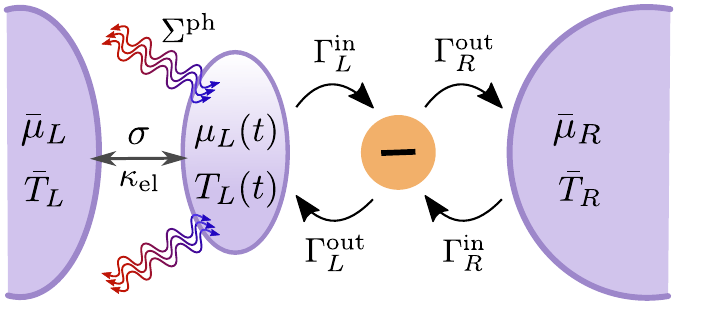}
    \caption{Sketch of the setup. A QD is coupled to a infinite reservoir on the right. On the left it is coupled to a finite reservoir in a quasi-equilibrium state described by time-dependent $\mu_L(t)$ and $T_L(t)$.     $\Gamma_\alpha^\text{in,out}$ with $\alpha = L,R$ are the rates for tunneling between the QD and the left (finite) and right (infinite) reservoirs.  The finite reservoir is in turn in contact with a  infinite one. The exchange between them is characterized by the electrical conductivity $\sigma$ and thermal conductivity  $\kappa_\text{el}$. The wavy lines indicate the phonon contribution $\Sigma^\text{ph}$ to the heat current due to electron-phonon interaction.}
    \label{fig:setup}
\end{figure}

The system consists of a single level, spin-degenerate QD tunnel-coupled to an infinite reservoir on the right and a finite reservoir on the left, see \cref{fig:setup}. The finite reservoir is additionally in contact with a second infinite reservoir to the left. All reservoirs are described by non-interacting electrons.   For the finite reservoir we assume that the internal relaxation is fast compared to any other time-scale in the system,  a common assumption  for bulk metals and semiconductors. Then,  it relaxes to the quasi-equilibrium state with a well-defined (but time-dependent) chemical potential  $\mu_L(t)$ and temperature $T_L(t)$, and is described by the Fermi-Dirac distribution $f(E, \mu_L(t), T_L(t)$ with  $f(E, \mu, T)~=~\{\exp\left[(E-\mu)/T\right]+1\}^{-1}$.  
Throughout the paper we set Boltzmann's constant $k_B = 1$, the elementary charge $e = 1$ and $\hbar = 1$.

A phenomenological rate equation to describe the transport through the QD system is given by
\begin{align}
    \frac{d}{dt}\begin{pmatrix} p_0 \\ p_1 \end{pmatrix} = \sum_\alpha
    \begin{pmatrix}
        -\Gamma^\text{in}_\alpha & \Gamma^\text{out}_\alpha \\
        \Gamma^\text{in}_\alpha & - \Gamma^\text{out}_\alpha 
    \end{pmatrix}
    \begin{pmatrix} p_0 \\ p_1 \end{pmatrix},
    \label{eq:QD_rate_eq}
\end{align}
where $p_0, p_1$ denote the probabilities of the QD orbital with energy $\epsilon$ being empty or occupied with either a spin-up or a spin-down electron, and $\alpha = L,R$ indexes the left and right reservoirs. At all times $p_0(t) + p_1(t) =1$ holds. \cref{eq:QD_rate_eq} is valid for a QD system without applied magnetic field, and in the limit of large Coulomb interaction where the state with two electrons on the QD can be neglected.
The rates for an incoming (or outgoing) electron with respect to the QD from (or to) the left, $\Gamma_L^\text{in}$ and $\Gamma_L^\text{out}$,  are time-dependent due to the finite nature of the reservoir. They are given by \cite{IngoldNazarov}
\begin{equation}
\begin{aligned}
    \Gamma^\text{in}_L(t) &= 2 \Gamma_L f\left(\epsilon, \mu_L(t), T_L(t)\right),\\
    \Gamma^\text{out}_L(t) &= \Gamma_L \left[1- f\left(\epsilon, \mu_L(t), T_L(t)\right)\right],
\label{eq:rates_L}
\end{aligned}
\end{equation}
where $\Gamma_L$ is the tunneling rate between the QD and finite reservoir and $\epsilon$ is the energy of the QD level. The total rate $\Gamma_L^\text{in} \sim 2 \Gamma_L$ in \cref{eq:rates_L} takes into account the spin degeneracy of the state with one electron on the QD.
For the  exchange with the right reservoir the tunnel rates are time-independent and given by
\begin{equation}
\begin{aligned}
    \Gamma^\text{in}_R &= 2 \Gamma_R f\left(\epsilon, \bar{\mu}_R, \bar{T}_R\right),\\
    \Gamma_R^\text{out} &= \Gamma_R \left[1- f\left(\epsilon, \bar{\mu}_R, \bar{T}_R\right)\right],
\label{eq:rates_R}
\end{aligned}
\end{equation}
where $\Gamma_R$ is the tunneling rate between the QD and infinite reservoir.
The description of the QD transport with the rate equation (\ref{eq:QD_rate_eq}) is applicable in the weak coupling limit where $\Gamma_\alpha \ll T_\alpha$.
Throughout, we denote the time-independent chemical potential and temperature of the infinite reservoirs with $\bar{\mu}_\alpha$ and $\bar{T}_\alpha$ with $\alpha = L,R$.

The electron current, $I_{L/R}^\text{QD}$, and energy current, $J_{L/R}^\text{QD}$, flowing out from the QD to the finite (L) and infinite (R) size reservoirs can be expressed in terms of the QD occupation probabilities and tunnel rates as
\begin{equation}
\begin{aligned}
I^\text{QD}_{L/R}(t) & =  p_1(t)\Gamma^\text{out}_{L/R}(t)-p_0(t)\Gamma^\text{in}_{L/R}(t), \\
J^\text{QD}_{L/R}(t) & = \epsilon I^\text{QD}_{L/R}(t).
\label{eq:QD_current}
\end{aligned}
\end{equation}
If $I_L^\text{QD} < 0$, the current flows out of the finite reservoir into the QD, while for $I_L^\text{QD} > 0$ it flows in the opposite direction.

Assuming the contact between the finite and infinite reservoirs on the left to have energy independent transport properties, the charge and energy currents flowing out of the infinite reservoir are given by \cite{Attard} 
\begin{equation}
\begin{aligned}
I_L(t)&=-\sigma \Delta \mu(t), \\ 
J_L(t)&=-\kappa_\text{el}\Delta T(t)+\frac{1}{2}\left(\mu_L(t)+\bar \mu_L\right)I_L(t).
\label{eq:lin_current}
\end{aligned}
\end{equation}
where we introduced $\Delta \mu(t)~=~\mu_L(t)-\bar \mu_L$ and $\Delta T(t)~=~T_L(t)-\bar T_L$. Here, $\sigma$ is the electrical conductance and we take the temperature-dependent electronic thermal conductance to be
\begin{align}
\kappa_\text{el}\left(\tilde T(t)\right)=\frac{\pi^2}{3}\sigma \tilde T(t),
\label{eq:kappa}
\end{align}
where $\tilde T(t)=(T_L(t)+\bar T_L)/2$.  In the limit where the difference in temperature of the left infinite and finite reservoir is negligible,  \cref{eq:kappa} reduces to the Wiedemann-Franz law
\cite{Kittel}.

To self-consistently determine the fluctuating $T_L(t)$ and $\mu_L(t)$ of the finite reservoir, we make use of the continuity, or conservation, equations for the charge and energy on the reservoir. The time derivative of the number of electrons of the reservoir, $N_L(t)$, is equal to the sum of the charge currents flowing into the reservoir
\begin{align}
\dot N_L(t)=I^\text{QD}_L(t)+I_L(t),
\label{eq:rate_charge}
\end{align}
where, within our free electron assumption for the finite reservoir, we have 
\begin{equation}
\begin{aligned}
\dot{N}_L(t)&= \nu_0 \frac{d}{dt}\int dE~f\left[E,\mu_L(t),T_L(t)\right]\\ 
&=\nu_0\dot{\mu}_L(t).
\label{eq:eom_charge}
\end{aligned}
\end{equation}
Here, $\nu_0$ is the density of states of the finite reservoir, which we take to be energy independent for simplicity, equal to the quantum capacitance of the reservoir. We note that, for simplicity, we neglect possible geometrical capacitive couplings (e.g., from the other reservoirs) to the excess electrons on the reservoir.

In the same way, the time derivative of the total energy of the finite reservoir is equal to the sum of the energy currents flowing into the reservoir
\begin{align}
\dot E_L(t)=J^\text{QD}_L(t)+J_L(t).
\label{eq:rate_energy}
\end{align}
where
\begin{equation}
\begin{aligned}
\dot{E}_L(t)&= \nu_0 \frac{d}{dt}\int dE E f\left[E,\mu_L(t),T_L(t)\right] \\
&= \nu_0 \mu_L(t) \dot{\mu}_L(t)+C_\text{el}[T_L(t)]\dot{T}_L(t),
\label{eq:eom_energy}
\end{aligned}
\end{equation}
where $C_\text{el}(T)=\nu_0\pi^2T/3$ is the specific heat of a free Fermi gas. Combining \cref{eq:rate_charge,eq:rate_energy,eq:eom_charge,eq:eom_energy} we can write
\begin{equation}
\begin{aligned}
&C_\text{el}[T_L(t)]\dot{T}_L(t)  \\
&= J^\text{QD}_L(t)+J_L(t)-\mu_L(t)\left[I^\text{QD}_L(t)+I_L(t)\right]  \\
&\equiv \dot{Q}^\text{QD}_L(t)+\dot{Q}_L(t).
\label{eq:eom_T}
\end{aligned}
\end{equation}
Here,  we introduced the heat currents of the finite reservoir
\begin{equation}
\begin{aligned}
\dot{Q}^\text{QD}_L(t)&= (\epsilon - \mu_L(t)) I_L^{\text{QD}}(t),  \\
\dot{Q}_L(t)&= J_L(t)-\mu_L(t)I_L(t),
\label{eq:heat_current}
\end{aligned}
\end{equation}
which are positive for heat flowing into the reservoir and negative for heat flowing out of it. 
This shows that the dynamics of the reservoir temperature $T_L(t)$ is governed by the heat flows into the reservoir, as one would expect. Taken together, \cref{eq:QD_rate_eq,eq:rate_charge,eq:eom_charge,eq:rate_energy,eq:eom_energy,eq:eom_T} provide a full description of the system dynamics for arbitrary applied bias voltages and temperatures $\bar{T}_\alpha\gg \Gamma_\alpha$.
In particular,  the evolution of $\mu_L(t)$ and $T_L(t)$ of the finite reservoir is governed by
\begin{equation}
\begin{aligned}
     \dot{\mu}_L(t) & = \frac{1}{\nu_0} \left[I^\textrm{QD}_L(t) + I_L(t)\right],\\
    \dot{T}_L(t) &= \frac{\dot{Q}_L^\textrm{QD}(t) + \dot{Q}_L(t)}{C_\text{el}\left(T_L(t)\right)}.
    \label{eq:meso_t}
\end{aligned}
\end{equation}

A strict linear response approach for a small applied bias reveals  two relevant dimensionless parameters governing the stationary transport properties, see  \cref{app:LinResponse} for details. The first one is $T\sigma/\Gamma_L$. The ratio $\sigma/\Gamma_L$ (in units of inverse temperature) characterizes the deviations from a system with only infinite reservoirs in the stationary state, which  is apparent from \cref{eq:QD_current,eq:lin_current}. The second parameter,  $\epsilon/T$,  describes how far from resonance the QD level is.

\section{Stationary state transport properties} \label{sec:transport}

In the following, we determine the transport properties of the set-up shown in \cref{fig:setup} in a non-equilibrium setting. A symmetrically applied bias voltage $V_b$ across the two infinite reservoirs sets their chemical potentials $\bar{\mu}_L = -\bar{\mu}_R = V_b/2$. The energy level $\epsilon$ of the QD can be tuned by an additional gate voltage $V_g = -\epsilon$ (for simplicity we set the gate lever arm to unity). Finally, we assume equal temperatures of the infinite reservoirs on the left and right,  $\bar{T}_L = \bar{T}_R = T$, as well as equal tunneling rates, $\Gamma_L = \Gamma_R = \Gamma$. 
We analyse the stationary electron current $I_L^{\text{QD,s}}~=~\lim_{t\to\infty} I_L^{\text{QD}}(t)$; we use the superscript s for the stationary state throughout.

To find the stationary state we solve the coupled equations of motions defined by \cref{eq:QD_rate_eq,eq:rate_charge,eq:eom_charge,eq:rate_energy,eq:eom_energy,eq:eom_T} for $\dot{\mu}_L = \dot{T}_L = \dot{p}_{0,1} = 0$. The  stationary state solution for the probabilities is  given by
\begin{equation}
\begin{aligned}
    p_0^s &= 1 - p_1^s, \\
    p_1^s &=  \frac{\sum_{\alpha} \Gamma_\alpha^{\text{in},s}}{\sum_{\alpha}\left(\Gamma_\alpha^{\text{in},s} + \Gamma_\alpha^{\text{out},s}\right)}.
\label{eq:p1_stationary}
\end{aligned}
\end{equation}
A general closed-form expression for the stationary state values for $\mu_L^s$ and $T_L^s$ of the finite reservoir is unattainable and we rely on a numerical solution for the transport properties.

\begin{figure}[t!]
\centering
\includegraphics[width = \linewidth]{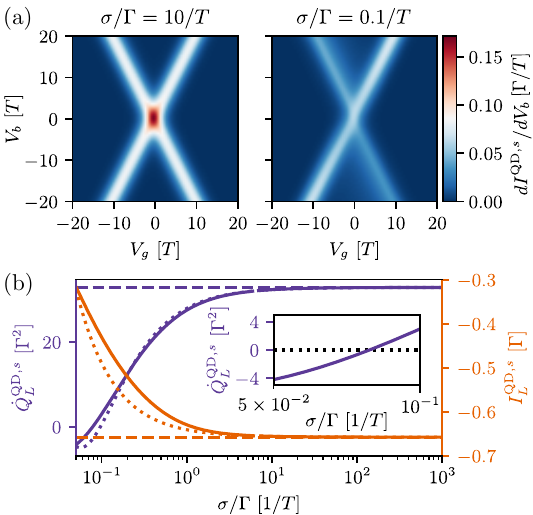}
\caption{ (a) Stability diagrams showing $dI^{\text{QD},s}_L/dV_b$ as a function of $V_b$ and $V_g$ for two different values of $\sigma/\Gamma$. For $\sigma/\Gamma = 10/T$ the transport looks qualitatively similar to a  single level QD coupled to infinite source and drain leads.  For $\sigma/\Gamma = 0.1/T $ the visible effects of the finite reservoir include an overall suppression of $I_L^{\text{QD},s}$ and a broadening and shift of one conductance peak. 
(b) Heat current $\dot{Q}_L^{\text{QD},s}/\Gamma^2$ (purple, left axis) and electron current $I_L^{\text{QD},s}/\Gamma$ (orange, right axis) as a function of $\sigma /\Gamma$ at fixed $V_b/T = 10$ and $V_g/T = 0$. The dotted lines show the analytical stationary state solution up to second order in $\Gamma/\sigma$ according to \cref{eq:PTGammaSigma}. For large $\sigma /\Gamma$ both $\dot{Q}_L^{\text{QD},s}$ and $I_L^{\text{QD},s}$ asymptotically approach the values for the QD coupled to infinite source and drain reservoirs, indicated by the dashed lines. For smaller $\sigma/\Gamma$ the stationary state values deviate, and can even lead to $\dot{Q}_L^{\text{QD},s} < 0$, shown in the inset. In all plots $\bar{T}_L = \bar{T}_R = T = 10 \Gamma$. }
\label{fig:stationary_transport}
\end{figure}

For the stationary current $I_L^\text{QD,s} =  \sigma \Delta \mu$ must hold due to current conservation. Therefore, the magnitude of the current is ultimately limited by $\sigma$,  if the QD level does not block the transport, i.e., if $\mu_L^s \gtrsim \epsilon  \gtrsim \bar{\mu}_R$ for $V_b > 0$.
Consequently, we can identify two effects of the finite reservoir on the stationary transport properties compared to coupling of the QD to infinite reservoirs: (i) A modification of the range where electrons can pass through the dot, and (ii) a suppression of the current due to finite $\sigma$.
There is no strict bound on the temperature $T_L^s$ as the heat current through the system is not conserved. Due to coupling of temperature and chemical potential, there is also no strict bound on $\mu_L^s$.
 Figure \ref{fig:stationary_transport}(a) shows the differential conductance $d I_L^{\text{QD},s}/d V_b$  for the QD system as a function of $V_g$ and $V_b$ for two different values of $\sigma/\Gamma$. For a single QD coupled to two infinite reservoirs, the stability diagram displays a  Coulomb diamond, or a pair of crossing conductance lines for infinite charging energy.
Considering now the finite reservoir in \cref{fig:stationary_transport}(a) with $\sigma/\Gamma = 10$, changes of $\mu_L^s$ in comparison to $\bar{\mu}_L$ are small and the qualitative features of the resulting stability diagram are already similar to the case $\sigma/\Gamma \to \infty$; this limit corresponds to effectively coupling the QD to two infinite reservoirs. 

For smaller values of $\sigma/\Gamma$ (focusing on $V_b>0$),  the conductance lines associated with tunneling between the QD and the finite reservoir are broadened and shifted to higher $\lvert V_b\rvert$. 
To understand this stationary state behavior it is useful to consider the full time dependence. If $\mu_L~=~\bar{\mu}_L$ at the initial time, electrons tunneling out of the finite reservoir and into the QD will lower $\mu_L$ until the current between the infinite and finite reservoirs (given by $\sigma \Delta \mu$) becomes equal to $I_L^\textrm{QD}$. For small $\sigma / \Gamma$, this happens for $\mu_L$ close to (or below) $\epsilon$. Thus, in the stationary state $\mu_L^s$ is pinned close to $\epsilon$ over a range in $V_b$ and the conductance peak shifts and broadens.

The width and  position of the conductance peaks associated with tunneling between the QD and right reservoir remain largely unaffected, especially for large $V_b, V_g$, but their heights are reduced as $I_L^{\text{QD},s}$ is overall suppressed for small $\sigma/\Gamma$.
If the QD would also be coupled to a finite reservoir on the right this peak would broaden and shift.
For $V_b < 0$ the explanations above hold for reversed inequalities. 

Further analytical insight can be gained in the regime of large $\sigma/\Gamma$ by writing
\begin{equation}
\begin{aligned}
    \mu_L(t) & = \bar{\mu}_L + \delta \mu_L(t), \\
    T_L(t) & = T + \delta T_L(t).
    \label{eq:small_delta}
\end{aligned}
\end{equation}
Assuming $\delta \mu_L, \delta T_L$ to be small we expand \cref{eq:QD_rate_eq,eq:meso_t} to first order in the small deviations. This is a good approximation when the finite reservoir is not (significantly) depleted because $\sigma/\Gamma \gg 1$.
It does not rely on a small applied bias voltage (see \cref{app:LinResponse} for the linear-response result for small bias voltage across the QD). The stationary state solutions for $\delta\mu_L^s$ and $\delta T_L^s$ are given in \cref{eq:deltaTLs} in \cref{app:EOMs}. 
 
In the limit $\sigma/\Gamma \to \infty$, $\mu_L^s\to\bar{\mu}_L$ and $T_L^s\to T$, and we recover the results of a QD coupled to two infinite reservoirs. This is exemplified by the stationary heat and electron currents shown in \cref{fig:stationary_transport}(b), including both the full numerical solution (solid lines) and  the analytical solution for small $\delta \mu_L^s, \delta T_L^s$ up to second order in $\Gamma/\sigma$ (dotted lines). 

We find good qualitative agreement of the analytical result with the numerics for the parameters in \cref{fig:stationary_transport}(b).  In general, $V_b$ and $\epsilon$ affect the qualitative agreement, and for small $V_b$ the analytics capture the behavior across the full range of $\sigma/\Gamma$. The expansion of the analytical results in \cref{eq:deltaTLs} to second order in $\Gamma/\sigma $ captures a suppression of the currents $\sim \Gamma/\sigma$ (see also \cref{eq:PTGammaSigma} in \cref{app:EOMs}) compared to the currents expected for the system if the QD was coupled to infinite reservoirs.
However, making $\sigma/\Gamma$ smaller than $\sim 1$,
eventually leads to violating the normalization of the probabilities $p_{0,1}$, 
 which implies a breakdown of the assumption of small deviations.

\section{Refrigeration of the finite reservoir} \label{sec:cooling}

Cooling takes place when the net heat current is negative, which is equivalent to heat flowing out of the finite reservoir.  The heat current consists of two contributions defined in \cref{eq:heat_current},   $\dot{Q}_L^{\text{QD},s}$ transported due to electrons tunneling into (out of) the QD and  $ \dot{Q}_L^{s}$  flowing between the finite and infinite reservoirs.   Here we investigate  the conditions necessary for refrigeration of the finite reservoir. 

For $V_b>0$ we have that  $I_L^{\text{QD},s}< 0$ and $\dot{Q}_L^{\text{QD},s}$ is negative when $\epsilon > \mu_L^s$. 
For $V_b< 0$ we instead have that  $I_L^{\text{QD},s}> 0$ and $\dot{Q}_L^{\text{QD},s}$ is negative  when $\epsilon < \mu_L^s$.  
The heat current $ \dot{Q}_L^{s}$ from the infinite reservoir will always positive if $T_L^s < T$.  Therefore,  the magnitude of  $\dot{Q}_L^{s}$  has to be  smaller than the heat current flowing out to the QD, i.e., $\lvert \dot{Q}_L^{s} \rvert  = \lvert - \kappa_\textrm{el} \Delta T^s + \sigma \left(\Delta \mu^s\right)^2/2\rvert < \lvert \dot{Q}_L^{\text{QD},s}\rvert$. If those requirements are fulfilled the net heat current is negative and a temperature of the finite reservoir $T_L^s ~<~T$ is induced.

 \begin{figure}[t!]
\centering
\includegraphics[width = \linewidth]{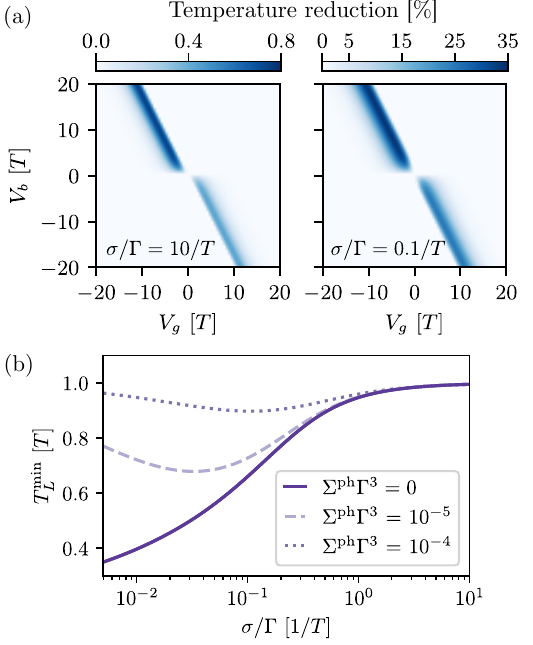}
\caption{ (a) Temperature reduction for the finite reservoir as a function of $V_b$ and $V_g$ for  $\sigma/\Gamma = 10/T$ (left) and $\sigma/\Gamma = 0.1/T$ (right). Plotted is the percentage $\lvert\Delta T^s\rvert/T~=~\lvert(T_L^s~-~T)\rvert~/~T$; regions with heating where $T_L^s > T$ are not shown.
(b)  $T_L^\text{min}$: $T_L^s$ minimized along cuts of fixed $V_b/T = 10$, plotted as a function of $\sigma /\Gamma$.  The dotted and dashed lines indicate $T_L^{\text{min}}$ including the electron-phonon interaction $\Sigma^\text{ph}$ in the thermal conductivity. All other parameters are chosen as in \cref{fig:stationary_transport}. }
\label{fig:minT}
\end{figure}

The parameter regime  where refrigeration is possible is shown in \cref{fig:minT}(a) for two different choices of $\sigma/\Gamma$ as a function of $V_b$ and $V_g$. 
First, the region in the $(V_g, V_b)$ plane where cooling of the finite reservoir is possible increases for smaller $\sigma/\Gamma$. For $V_b>0$, this correlates with a larger region where $\epsilon~>~\mu_L^s$ is fulfilled due to the larger shift of  $\mu_L^s$ compared to larger $\sigma/\Gamma$. Similarly, for $V_b~<~0$ we have instead $\epsilon < \mu_L^s$. 
Second, the difference between $T_L^s$ and $T$ increases (non-linearly) with decreasing $\sigma/\Gamma$, showing a reduced temperature down to $T_L^s = 0.65 T$ for $\sigma/\Gamma = 0.1$. 
The slight bias asymmetry in the temperature reduction is related to spin degeneracy, which causes an asymmetry between the rates for electrons tunneling in and out of the QD.

A key question for any cooling application is what minimum temperature  $T_L^{\text{min}}$ of the finite reservoir can be reached. We  determine  $T_L^{\text{min}}$ across a line cut at constant $V_b$ (the results only depends very weakly on the choice as long as $V_b\gg T$). 
The $T_L^{\text{min}}$ that can be reached depending on $\sigma$ is shown in \cref{fig:minT}(b). 
It keeps decreasing with decreasing $\sigma/\Gamma$, but, as we will see in \cref{sec:transient}, the time needed to reach the stationary state increases because the rate of cooling decreases. This also means that the cooling is more sensitive to other sources of heat flows for small $\sigma/\Gamma$.

The phonon contribution to the heat exchange is expected to play an important role in quantum devices \cite{Cahill2014,Qian2021}. Assuming that the phonon temperature is given by $T_{p} =T$, the heat current flowing from the finite reservoir into the phonon bath can be included in \cref{eq:heat_current} through a modified thermal conductivity, where we replace
\begin{align}
    \kappa_\text{el} \to \kappa_\text{el} +  \Sigma^\text{ph} T^4,
\end{align}
with the electron-phonon coupling strength $\Sigma^\text{ph}$. For the temperature dependence we choose the typical $T^4$ for metals \cite{Pekola2018,Brange2018}, however, the exact dependence is not crucial and different exponents are observed experimentally \cite{Giazotto2006}.
The  phonon contribution becomes more prominent for small $\sigma/\Gamma$, leading to a non-monotonic behaviour of $T_L^{\text{min}}$ as a function of $\sigma/\Gamma$ with a clear minimum. The reason is that for small $\sigma/\Gamma$ the electron current is lower which corresponds to a lower heat current out of the finite reservoir. This leads to a lower cooling power and therefore a larger impact of the phonons.  
An additional source of heat exchange in this system, with the potential to influence $T_L^{\text{min}}$ further, are higher-order tunneling processes  \cite{Turek2002}. Within the weak coupling regime ($\Gamma \ll T$) studied here, such contributions are small.

\section{Full time evolution} \label{sec:transient}

We now investigate the full system dynamics to gain insight into the typical timescales, including the time needed to cool the finite reservoir. 

\begin{figure}[t!]
\centering
\includegraphics[width = \linewidth]{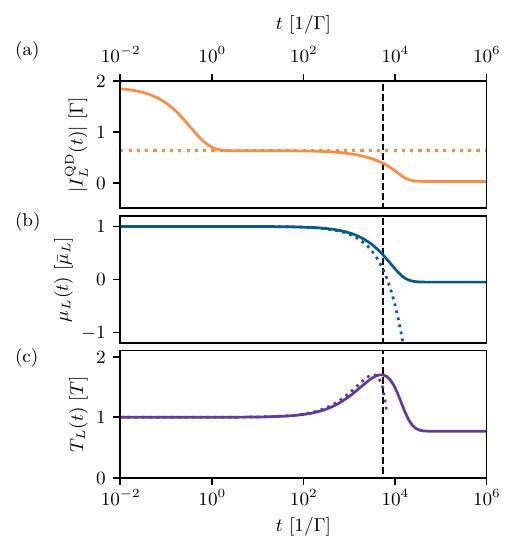}
\caption{Full time evolution of (a) $\lvert I_L^\text{QD}(t)\rvert$, (b) $\mu_L(t)$ and (c) $T_L(t)$ for the initial condition  $\mu_L(0) = \bar{\mu}_L$,  $T_L(0)~=~\bar{T}_L$, $p_0(0)~=~1$, $V_b = 10 T$,  $T~=~ 10 \Gamma$, $\epsilon~=~V_b/4$,  $\sigma/\Gamma~=~5~\cdot~10^{-3}/T$ and $\nu_0=1000/T$. The black dashed line marks the time when $\mu_L(t)$ = $\epsilon$ . 
The dotted line in (a) indicates $I_L^{\text{QD},\textrm{inf}}$ and the dotted lines in (b) and (c) indicate the evolution according to \cref{eq:meso_short_t}. } 
\label{fig:transient}
\end{figure}

Beyond the stationary state we have access to the full time evolution by numerically integrating the 
 coupled equations of motions for ${p}_{0,1}(t),\,{\mu}_L(t)$ and ${T}_L(t)$ defined in \cref{eq:QD_rate_eq,eq:meso_t}. For further physical insight we focus on a distinct regime where the QD dynamics sets the fastest time scale. We assume $\sigma/ \Gamma \ll 1 $ and $V_b \gg T$, and we consider an initial state where the finite reservoir is in equilibrium with the left infinite reservoir, and the QD is unoccupied, $p_0(t=0) = 1$.  Additionally, we choose $\epsilon/T = V_b/4$, such that the finite reservoir is cooled in the stationary state. The transients for $I_L^\text{QD}(t),\ \mu_L(t)$ and $T_L(t)$ are shown in \cref{fig:transient}.

Initially, the fast QD dynamics on the time scale of $\sim~1/\Gamma$ is 
associated with establishing the QD occupation  $p_0^{\textrm{inf}}$ and current $I^{\text{QD},\textrm{inf}}_L$ that would be the stationary state values if the QD was coupled to an infinite reservoir. The initial dynamics of $I_L^\textrm{QD}(t)$ is largely independent of $\mu_L(t)$ as long as $\mu_L(t) >\epsilon$ because the Fermi function determining the electron tunneling rate is close to one. 
 In response, $\mu_L(t)$ and $T_L(t)$ adjust  for intermediate times $t > 1/\Gamma$ according to 
\begin{equation}
 \begin{aligned}
     \dot{\mu}_L(t) &\sim \frac{1}{\nu_0}I^{\text{QD},\textrm{inf}}_L, \\
      \dot{T}_L(t)  & \sim  \frac{\left[\epsilon - \mu_L(t)\right] }{C_\text{el}\left[T_L(t)\right]} I_L^{\text{QD},\textrm{inf}},
\end{aligned}
\label{eq:meso_short_t}
\end{equation}
which follows from \cref{eq:meso_t} with  $\Delta \mu_L, \Delta T_L \ll 1$ and 
\begin{align}
    I^{\text{QD},\textrm{inf}}_L \sim - 2 \Gamma   p_0^{\textrm{inf}}.
 \end{align}
Consequently, the initial downward shift of $\mu_L(t)$ is linear, while $T_L(t)$ first increases
as only electrons at energy $\epsilon$ can tunnel into the QD, and electrons from below $\mu_L(t)$ are removed from the finite reservoir.
As soon as $\mu_L(t) < \epsilon$, $\dot{Q}_L^\textrm{QD}$ becomes negative and  the finite reservoir eventually starts to cool  (although $\dot{Q}_L$ remains positive). Simultaneously, the current decreases since the tunneling rate is sensitive to the tail of the Fermi function.  At these later times, 
the feedback between $\mu_L(t), T_L(t)$ and $p_j(t)$ via the Fermi function $f(\epsilon,\mu_L(t),T_L(t))$ becomes important.  
  We can estimate the longest possible time scale of the system, which is predominately set by $\sigma/\nu_0$ as the exponential evolution according to $\dot{\mu}_L \sim \sigma\Delta \mu(t)/\nu_0$ and $\dot{T}_L \sim -\sigma(T_L^2(t) - T^2)/C_\textrm{el}(T_L(t))$ from \cref{eq:meso_t} set in.
  However, the exact details of how the stationary state is reached depends strongly on the chosen system parameters, as well as the initial condition.

\section{Conclusion}

In this work, we investigated the non-equilibrium transport properties of a QD  coupled to a finite reservoir, as well as two infinite reservoirs. 
The finite nature of the reservoir leads to (i) a shift of the chemical potential of the finite reservoir which in turn leads to  a broadening of one of the conductance peaks in the charge stability diagram; (ii) a suppressed electron current compared to the current that would flow in the case of coupling to infinite reservoirs. 
Along the conductance peaks we find a region where a temperature $T_L^s$ smaller than the temperature of the infinite reservoirs can be established. Therefore, the electron transport and associated heat flow through a finite reservoir leads to stationary-state refrigeration of that reservoir. 
This result is immediately relevant for refrigeration applications, as any physical system to be cooled is necessarily finite.
The minimum temperature $T_L^\text{min}$ that can be reached strongly depends on the coupling to the infinite reservoir and is subject to a trade-off.
Decreasing $\sigma/\Gamma$ results in decreasing  $T_L^\text{min}$, but  also leads to an increasing shift of the chemical potential and time needed to reach the stationary state. Coupling to phonons strongly impacts $T_L^\text{min}$ for small $\sigma/\Gamma$, and leads to a minimum of $T_L^\text{min}$ for finite $\sigma/\Gamma$.
Furthermore, we studied the transient dynamics of the system and identified three distinct regimes. An important timescale is set by the ratio of the electrical conductivity with the density of states of the finite reservoir, $\sigma/\nu_0$. Therefore, a choice of small $\sigma/\Gamma$ for a minimal temperature also leads to slow dynamics towards the stationary state.

Utilizing microscopic approaches \cite{Riera2021} for finite reservoirs might provide a promising route to 
 address more complicated quantum systems to further study non-equilibrium transport including fluctuations and noise. But also more fundamental questions such as the modification of the energy distribution \cite{Ajisaka2013}, especially in quasi-equilibrium or even out-of-equilibrium reservoir states \cite{Trushechkin2022} could be addressed.

%

\begin{acknowledgments}
We would like to thank Viktor Svensson and Konstantin Nestmann for insightful discussions. 
We acknowledge financial support from the Wallenberg Center for Quantum Technologies (WACQT), the Swedish Research Council (Grant Agreement No.\,2020-03412 and No.\,2018-03921), and from NanoLund. SM acknowledges financial support from the PNRR-MUR project and PE0000023-NQSTI project, co-funded by the European Union -- NextGeneration EU, and from Provincia Autonoma di Trento (PAT).
SVM acknowleges funding from the Knut and Alice Wallenberg Foundation (Project No.\,2016-0089). This project is co-funded by the European
Union (Quantum Flagship project ASPECTS, Grant Agreement No.\,101080167). Views and opinions expressed are
however those of the authors only and do not necessarily
reflect those of the European Union, Research Executive
Agency or UKRI. Neither the European Union nor UKRI can
be held responsible for them. 

\end{acknowledgments}

%

\appendix
\section{Linear response theory for stationary state properties} \label{app:LinResponse}

We consider the stationary electron and energy currents for a small, symmetrically applied bias voltage $\Delta V_b$ across the infinite reservoirs, such that $\bar{\mu}_L = -\bar{\mu}_R~=~\Delta V_b/2$. There is no temperature gradient across the system, $\bar{T}_L =~\bar{T}_R~=~T$ and we assume equal tunneling rate $\Gamma_L = \Gamma_R = \Gamma$. The chemical potential and temperature of the finite reservoir are
\begin{equation}
\begin{aligned}
    \mu_L & = \delta \mu_L^s\\
    T_L & = T + \delta T_L^s.
\end{aligned}
\end{equation}
In the strict linear response regime, the quantities $\delta \mu_L^s$ and $\delta T_L^s$ are proportional to the applied bias $\Delta V_b$. 
First, the charge and energy currents between the left infinite and finite reservoirs are given by
\begin{equation}
\begin{aligned}
    I_{L} & = -\sigma\left( \delta \mu_L^s -\frac{\Delta V_b}{2}\right) \\
    J_{L} & = -\kappa_\text{el} \delta T_L^s.
\end{aligned}
\end{equation}
Second, the particle and energy currents between the QD and the right infinite reservoir are
\begin{equation}
    \begin{aligned}
        I_{R}^\textrm{QD}  & = p_1^s \Gamma^\text{out}_R - p_0^s\Gamma^\text{in}_R, \\
        J_{R}^\textrm{QD} &= \epsilon  I_{R}^\textrm{QD}.
    \end{aligned}
\end{equation}
In the stationary state, $I_L = I_R^\textrm{QD}$ and $J_L = J_R^\textrm{QD}$ are required. Solving for $\delta \mu_L^s$ and $\delta T_L^s$ leads to
\begin{equation}
    \begin{aligned}
        \delta\mu_L^s & = \frac{\Delta V_b}{2} \frac{\frac{\sigma}{G} + 3\left(\frac{\epsilon}{\pi T}\right)^2 - 1}{\frac{\sigma}{G} + 3\left(\frac{\epsilon}{\pi T}\right)^2 + 1}, \\
        \delta T_L^s & = - \Delta V_b \frac{3\frac{\epsilon}{\pi^2 T}}{\frac{\sigma}{G} + 3\left(\frac{\epsilon}{\pi T}\right)^2+1},
    \end{aligned}
\end{equation}
where we used the Wiedemann-Franz law $\kappa_\text{el} = \pi^2 \sigma T /3$ and introduced the conductance 
\begin{align}
    G = \frac{\Gamma}{T} \frac{1}{3\left[1+ \cosh{\left(\epsilon /T\right)}\right] - \sinh{\left(\epsilon/T\right)}}.
\end{align}
Here, we can identify two independent, dimensionless parameters, $\epsilon/T$ and $T\sigma/\Gamma$. The position of the QD level $\epsilon$ is crucial as  the electron current is exponentially suppressed for $\lvert\epsilon -\bar{\mu}_\alpha\rvert > T$. Additionally, $\sigma/\Gamma$ gives a measure of the competing couplings  $\sigma$ (between the infinite and the finite reservoir) and $\Gamma$ (between finite reservoir and QD).

\onecolumngrid
\section{Large coupling to the infinite reservoir} \label{app:EOMs}

In the limit of large coupling to the infinite reservoir, $T \sigma/\Gamma \gg 1$, the finite reservoir is barely depleted. We can assume that $\delta \mu_L$ and $\delta T_L$ in \cref{eq:small_delta} are small, i.e., $\delta \mu_L \ll \bar{\mu}_L$ and $\delta T_L \ll T$, and we can expand the Fermi functions in the equations of motion (see \cref{eq:QD_rate_eq,eq:meso_t}). The expansion goes beyond a linear response description as it is not limited to small applied bias. To first order in $\delta\mu_L,\,\delta T_L$ we find closed form expressions in the stationary state, which are given by
\begin{equation}
\begin{aligned}
    \delta \mu_L^s & = - \frac{\Gamma}{\sigma}\frac{2 \bar{f}_- \bar{T}_L \left[ 6\frac{\Gamma}{\sigma} \bar{f}_- \left(\epsilon - \bar{\mu}_L\right) - \pi^2 \left(2 + \bar{f}_+ \right) \bar{T}_L^2\right]}{\left(2 + \bar{f}_+ \right)\left[6 \frac{\Gamma}{\sigma} \bar{f}_- \left(\epsilon - \bar{\mu}_L\right) - \pi^2 \left(2 + \bar{f}_+ \right) \bar{T}_L^2\right]\bar{T}_L - 4 \frac{\Gamma}{\sigma} \left( 1 +\bar{f}_R\right) F_L \left[\pi^2 \bar{T}_L^2 + 3 \left(\epsilon - \bar{\mu}_L\right)^2\right]}, \\
    \delta T_L^s & = \frac{\Gamma}{\sigma} \frac{6\bar{f}_- \bar{T}_L^2\left[2 \frac{\Gamma}{\sigma}\bar{f}_- + \left(2 + \bar{f}_+\right)\left(\epsilon - \bar{\mu}_L\right)\right]}{\left(2 + \bar{f}_+ \right)\left[6 \frac{\Gamma}{\sigma} \bar{f}_- \left(\epsilon - \bar{\mu}_L\right) - \pi^2 \left(2 + \bar{f}_+ \right) \bar{T}_L^2\right]\bar{T}_L - 4 \frac{\Gamma}{\sigma} \left( 1 +\bar{f}_R\right) F_L \left[\pi^2 \bar{T}_L^2 + 3 \left(\epsilon - \bar{\mu}_L\right)^2\right]}.
    %
    \label{eq:deltaTLs}
\end{aligned}
\end{equation}
Here, we defined 
\begin{align}
    \bar{f}_- & = \bar{f}_L - \bar{f}_R,\\
    \bar{f}_+ & = \bar{f}_L + \bar{f}_R,\\
    F_L & = \bar{f}_L\left[1 - \bar{f}_L\right],
\end{align}
with $\bar{f}_\alpha = f(\epsilon, \bar{\mu}_\alpha, \bar{T}_\alpha)$. The stationary solution for the dot occupation probability according to \cref{eq:p1_stationary} within the small-deviations expansion is given by
\begin{align}
    p_1^s = \bar{p}_1^s + \frac{4 F_L}{\left(2 + \bar{f}_+\right)^2}\left[\frac{\delta\mu_L^s}{\bar{T}_L} + \frac{\epsilon -\bar{\mu}_L}{\bar{T}_L^2} \delta T_L^s\right],
\end{align}
with $\bar{p}_1^s = 2\bar{f}_+/(2 + \bar{f}_+)$.

The stationary state solution for the  small deviations from the infinite reservoir quantities in \cref{eq:deltaTLs} depends explicitly on the ratio $\Gamma/\sigma$, and in the limit of large $\sigma$, $\delta\mu_L^s, \delta T_L^s \to 0$. Expanding  \cref{eq:deltaTLs} in the ratio $\Gamma/\sigma$ up to second order leads to
\begin{equation}
    \begin{aligned}
        \delta \mu_L^s \approx & - \frac{2 \bar{f}_-}{2 + \bar{f}_+}  \frac{\Gamma}{\sigma}
         +  \frac{8 \bar{f}_-\left(1 + \bar{f}_R\right) F_L \left[\pi^2 \bar{T}_L^2 + 3 \left(\epsilon - \bar{\mu}_L\right)^2\right]}{\left(2 + \bar{f}_+\right)^3 \pi^2 \bar{T}_L^3} \left(\frac{\Gamma}{\sigma}\right)^2,\\
        \delta T_L^s \approx & - \frac{6 \bar{f}_- \left(\epsilon -\bar{\mu}_L\right)}{\left(2 + \bar{f}_+\right) \pi^2 \bar{T}_L} 
        \frac{\Gamma}{\sigma}
         - \frac{12 \bar{f}_- \left[ \bar{f}_- \left(2 +\bar{f}_+\right) \bar{T}_L - 2 \left(1 +\bar{f}_R\right) F_L \left(\epsilon - \bar{\mu}_L\right)\right]\left[\pi^2 \bar{T}_L^2 + 3 \left(\epsilon - \bar{\mu}_L\right)^2\right]}{\left(2 +\bar{f}_+\right)^3 \pi^4 \bar{T}_L^4} \left(\frac{\Gamma}{\sigma}\right)^2.
    \end{aligned}
    \label{eq:PTGammaSigma}
\end{equation}
These solutions are used to calculate the analytical stationary charge and heat current in \cref{fig:stationary_transport}(b) [dotted lines].

\twocolumngrid


\bibliography{references}

\end{document}